\documentclass[aps, prd, twocolumn, nofootinbib,]{revtex4-1}

\usepackage[top=1in, bottom=1in, left=1in, right=1in]{geometry}

\usepackage{amsmath}
\usepackage{amsfonts}
\usepackage{wasysym}
\usepackage{graphicx}
\usepackage{comment}
\usepackage{tensor}

\def\filetype{pdf}

\def\path{}

\allowdisplaybreaks[1]

\begin{document}


\title{Radial oscillations and stability of multiple-fluid compact stars}
\author{Ben Kain}
\affiliation{Department of Physics, College of the Holy Cross, Worcester, Massachusetts 01610, USA}

\begin{abstract}
\noindent I derive a system of pulsation equations for compact stars made up of an arbitrary number of perfect fluids that can be used to study radial oscillations and stability with respect to small perturbations.  I assume spherical symmetry and that the only inter-fluid interactions are gravitational.  My derivation is in line with Chandrasekhar's original derivation for the pulsation equation of a single-fluid compact star and keeps the contributions from the individual fluids manifest.  I illustrate solutions to the system of pulsations equations with one-, two-, and three-fluid examples.
\end{abstract} 

\maketitle


\section{Introduction}

The perfect fluid model has been very successful in modeling the matter content of compact stars \cite{GlendenningBook}.  This is particularly the case with fermionic matter, because the perfect fluid model is able to capture properties of the quantization of fermions, such as Pauli exclusion.

Static compact star solutions using perfect fluids are found by solving the time-independent Einstein field equations with a perfect fluid energy-momentum tensor.  Recently there has been increased interest in compact stars made with multiple fluids.  This has been motivated by the possibility that neutron stars might capture dark matter \cite{Kouvaris:2007ay, Bertone:2007ae, deLavallaz:2010wp}.  If dark matter cannot self-annihilate (as is the case, for example, with asymmetric dark matter \cite{Kaplan:2009ag, Zurek:2013wia}), it will accumulate in the neutron star.  In such a scenario, dark matter can be modeled as an additional fluid in the star.  There have been a number of recent studies of the properties of static two-fluid compact stars \cite{Sandin:2008db, Ciarcelluti:2010ji, Leung:2011zz, Leung:2012vea, Li:2012ii, Leung:2013pra, Goldman:2013qla, Xiang:2013xwa, Tolos:2015qra, Mukhopadhyay:2015xhs, Gresham:2018rqo, Deliyergiyev:2019vti}.

After a compact star solution is found, one can study the effect of small perturbations to the solution.  This may be done by solving a pulsation equation, which is a differential equation describing the time-evolution of the perturbation.  For spherically symmetric static solutions and perturbations, the solution to the pulsation equation gives the squared radial oscillation frequency.  In addition to describing radial oscillations, the squared radial oscillation frequency tells us about the stability of the star with respect to small perturbations.  If the squared oscillation frequency is negative, the compact star solution is unstable.

Chandrasekhar was the first to derive a pulsation equation for a compact star made of a single perfect fluid \cite{Chandrasekhar:1964zz}.  His pulsation equation was subsequently rewritten in various ways \cite{Misner:1974qy, Chanmugam, Gondek:1997fd, Kokkotas:2000up}, in some cases to facilitate numerical solutions.  More recently, a pulsation equation was derived for two-fluid systems  \cite{Comer:1999rs, Andersson:2006nr} and has been used to study the stability of two-fluid stars \cite{Leung:2011zz, Leung:2012vea, Leung:2013pra} (an alternative approach for determining the stability of two-component stars, which has been applied to boson-fermion stars \cite{Henriques:1989ar, Henriques:1989ez}, was studied in \cite{Henriques:1990xg, ValdezAlvarado:2012xc}).  In this construction, there is a single pulsation equation.  It is possible to solve the single pulsation equation in a two-fluid system, but without additional assumptions it can be difficult to do so.

One such assumption is to disallow nongravitational inter-fluid interactions.  Indeed, this assumption is made in many of the studies of two-fluid compact stars \cite{Sandin:2008db, Ciarcelluti:2010ji, Leung:2011zz, Leung:2012vea, Li:2012ii, Leung:2013pra, Goldman:2013qla, Xiang:2013xwa, Tolos:2015qra, Mukhopadhyay:2015xhs, Deliyergiyev:2019vti} (a notable exception is \cite{Gresham:2018rqo}).   This assumption allows the equations of motion to separate and leads to a system of coupled pulsation equations, with the same number of pulsation equations as fluids.  This increase in the number of pulsation equations makes them easier to solve.

In this work, I present this system of pulsation equations.  I shall assume spherical symmetry and, as mentioned, allow only gravitational inter-fluid interactions (I make no assumptions about self-interactions).  In doing this, I give a very different derivation than in \cite{Comer:1999rs, Andersson:2006nr}, a derivation made possible by my assumption of there only being gravitational inter-fluid interactions and a derivation in line with Chandrasekhar's original derivation for a single fluid.  The system of equations I arrive at keeps the contributions from the individual fluids manifest and allows them to be easily solved for, if desired.  I illustrate the use of this system of pulsation equations with one-, two-, and three-fluid examples.  

In the next section I derive the system of pulsation equations.  In Sec.\ \ref{sec:examples}, I present one-, two-, and three-fluid examples.  I conclude in Sec.\ \ref{sec:conclusion}.


\section{Equations}

In this section I derive a number of equations.  This includes the Tolman–Oppenheimer–Volkoff (TOV) equations adapted to a multi-fluid system.  Solutions to the TOV equations describe static compact stars and I will often refer to such solutions as equilibrium solutions.  I also derive the system of pulsation equations, which describe radial oscillations about the equilibrium solutions.

I consider only spherically symmetric spacetimes.  In deriving equations---in particular, the pulsation equations---it is convenient to use a spherically symmetric metric of the form
\begin{equation} \label{nu lambda metric}
ds^2 = -e^{\nu(t,r)} dt^2 + e^{\lambda(t,r)} dr^2 + r^2 d\Omega^2,
\end{equation}
where $d\Omega^2 = d\theta^2 + \sin^2\theta \,d\phi^2$.  I  note that later I will use a different form for the metric that is better suited for numerically solving the TOV and pulsation equations.

I assume there are an arbitrary number of perfect fluids, whose individual energy-momentum tensors separate:
\begin{equation} \label{T tot}
T_\text{tot}^{\mu\nu} = \sum_x  T_x^{\mu\nu},
\end{equation}
where $x$ labels the fluid,
\begin{equation} \label{Tx}
T_x^{\mu\nu} = (\epsilon_x + p_x)u_x^\mu u_x^\nu + p_x g^{\mu\nu}
\end{equation}
is the standard energy-momentum tensor for a perfect fluid, $u_x^\mu$ is the four-velocity of the fluid, and $\epsilon_x$ and $p_x$ can be thought of as the fluid's energy density and pressure.
$T_\text{tot}^{\mu\nu}$ in Eq.\ (\ref{T tot}) is the (total) energy-momentum tensor of the system and is what goes on the right hand side of the Einstein field equations:
\begin{equation} \label{GR eq}
G^{\mu\nu} = 8\pi G T_\text{tot}^{\mu\nu}.
\end{equation}
In addition to its energy-momentum tensor, the matter sector is defined by equations of state, which I also assume separate:
\begin{equation}
p_x = p_x (\epsilon_x),
\end{equation}
i.e.\ $p_x$ depends only on its associated energy density $\epsilon_x$ and not on any $\epsilon_{y\neq x}$.  The assumption that $T_x^{\mu\nu}$ and $p_x$ only depend on the variables of fluid $x$ means that there are only gravitational inter-fluid interactions.  This assumption greatly facilitates solving both the equilibrium and pulsation equations, in part because it means that the individual $T_x^{\mu\nu}$ are conserved,
\begin{equation} \label{nabla Tx}
\nabla_\mu T_x^{\mu\nu} = 0,
\end{equation}
in addition to the requisite $\nabla_\mu T_\text{tot}^{\mu\nu} = 0$.

In a spherically symmetric system, fluids cannot flow in the $\theta$- or $\phi$-directions and thus $u^\theta = u^\phi = 0$.  Following Chandrasekhar \cite{Chandrasekhar:1964zz}, I define $v_x \equiv e^{\nu/2} u_x^r$.  Using that $g_{\mu\nu} u_x^\mu u_x^\nu = -1$, I have $u_x^t = \exp(-\nu/2) \sqrt{1 + \exp(\lambda - \nu) v_x^2}$.  It is now straightforward to write the components of $T_x^{\mu\nu}$ in Eq.\ (\ref{Tx}) in terms of $v_x$.  Doing so is not particularly illuminating, nor will I be using $T_x^{\mu\nu}$ in such a form.  Instead, I write the system variables as perturbations about their equilibrium values:
\begin{align}
\nu(t,r) &= \nu_{0}(r) + \delta \nu(x,t)
\\
\lambda(t,r) &= \lambda_{0}(r) + \delta \lambda(x,t)
\\
\epsilon_x(t,r) &= \epsilon_{x0}(r) + \delta \epsilon_x(x,t)
\\
p_x(t,r) &= p_{x0}(r) + \delta p_x(x,t).
\end{align}
The equilibrium values are denoted by a subscripted $0$ and describe equilibrium, or static, solutions and hence are time-independent.  I note that the equilibrium value of $v_x$ vanishes and $v_x$ is itself at the order of a perturbation.  My interest is in using the perturbations to describe radial oscillations about the equilibrium solutions.  I therefore consider equations only to first order in the perturbations.  To this order, the nonzero components of $T_x^{\mu\nu}$ are
\begin{align}
(T_x)\indices{^t_t} 
&= - \epsilon_{x0} - \delta \epsilon_x
\\
(T_x)\indices{^t_r}
&= e^{\lambda_0-\nu_0}(\epsilon_{x0} + p_{x0})  v_x
\\
(T_x)\indices{^r_t} 
&= -(\epsilon_{x0} + p_{x0})  v_x
\\
(T_x)\indices{^r_r} &=
(T_x)\indices{^\theta_\theta} =
(T_x)\indices{^\phi_\phi} 
= p_{x0} + \delta p_x.
\end{align}
Summing these components over $x$ gives $(T_\text{tot})\indices{^\mu_\nu}$.  Setting the perturbations to zero, the equilibrium energy-momentum tensor is
\begin{equation}
(T_{x0})\indices{^\mu_\nu} = \text{diag}
(-\epsilon_{x0}, p_{x0}, p_{x0}, p_{x0}).
\end{equation}
Summing these components over $x$ gives $(T^\text{tot}_0)\indices{^\mu_\nu}$.  For the equilibrium solutions, we see that $\epsilon_{x0}$ and $p_{x0}$ are fluid $x$'s contributions to the (total) energy density and pressure of the system:
\begin{equation}
\epsilon^\text{tot}_0 = \sum_x \epsilon_{x0}, \qquad
p^\text{tot}_0 = \sum_x p_{x0}.
\end{equation}

The Einstein field equations in (\ref{GR eq}) lead to a number of equations that are written in terms of the components of the energy-momentum tensor and determine the metric functions $\nu$ and $\lambda$.  Three such equations are \cite{Chandrasekhar:1964zz,BaumgarteBook}
\begin{align}  
\nu' &= +8\pi G r e^{\lambda}  (T_\text{tot})\indices{^r_r} + \frac{e^{\lambda}-1}{r}
\label{metric eqs a}
\\
\lambda' &= 
-8\pi G r e^\lambda  (T_\text{tot})\indices{^t_t}
- \frac{e^\lambda - 1}{r} 
\label{metric eqs b}
\\
\dot{\lambda} &= -8\pi G r e^\nu  (T_\text{tot})\indices{^t_r},
\label{metric eqs c}
\end{align}
where a prime denotes an $r$-derivative and a dot denotes a $t$-derivative.  The equilibrium version of these equations are straightforward to write down:
\begin{align} 
\nu_0' &= 8\pi G r e^{\lambda_0} p_0^\text{tot} + \frac{e^{\lambda_0}-1}{r}
\label{nu lambda equil metric eqs a}
\\
\lambda_0' &= 
8\pi G r e^{\lambda_0}  \epsilon_0^\text{tot}
- \frac{e^{\lambda_0} - 1}{r},
\label{nu lambda equil metric eqs b}
\end{align}
where the equilibrium version of Eq.\ (\ref{metric eqs c}) vanishes identically.  The perturbed version of Eqs.\ (\ref{metric eqs a})--(\ref{metric eqs c}), after using Eqs.\ (\ref{nu lambda equil metric eqs a}) and (\ref{nu lambda equil metric eqs b}) to cancel the equilibrium parts, are
\begin{align} 
\delta \nu' &= 8\pi G r e ^{\lambda_0} ( p_0^\text{tot} \delta \lambda + \delta p^\text{tot})
+ \frac{e^{\lambda_0}}{r} \delta \lambda
\label{metric perturb a}
\\
\delta \lambda' &= 8\pi G r e ^{\lambda_0} (\epsilon_0^\text{tot} \delta \lambda + \delta \epsilon^\text{tot})
- \frac{e^{\lambda_0}}{r} \delta \lambda
\label{metric perturbb}
\\
\delta\dot{\lambda} &= -8\pi G r e^{\lambda_0} \sum_x (\epsilon^x_0 + p^x_0) v_x.
\label{metric perturb c}
\end{align}
There are two more equations for the metric functions that I will need.  The first is obtained by combining Eqs.\ (\ref{nu lambda equil metric eqs a}) and (\ref{metric perturb a}):
\begin{equation} \label{nu eq}
\delta \nu' - \nu_0' \delta \lambda = \frac{\delta \lambda}{r} + 8\pi G r e^{\lambda_0} \delta p^\text{tot}.
\end{equation}
The second is less commonly used than Eqs.\ (\ref{metric eqs a})--(\ref{metric eqs c}), but like them follows from the Einstein field equations.  I only need its equilibrium version \cite{Chandrasekhar:1964zz,BaumgarteBook}:
\begin{equation} \label{nu pp eq}
16\pi G e^{\lambda_0} p_0^\text{tot}
= \nu_0'' + \frac{1}{2} (\nu_0')^2
- \frac{1}{2} \nu_0' \lambda_0'
+ \frac{1}{r}(\nu_0' - \lambda_0').
\end{equation}

Lastly, I need the equations of motion.  I will obtain these from $\nabla_\mu T_x^{\mu\nu} = 0$, i.e.\ from conservation of the individual $T_x^{\mu\nu}$.  Using the metric in Eq.\ (\ref{nu lambda metric}) to evaluate the divergence, I find
\begin{align} 
0 &= \partial_t (T_x)\indices{^t_t} + \partial_r (T_x)\indices{^r_t} + \frac{\dot{\lambda}}{2} [(T_x)\indices{^t_t} - (T_x)\indices{^r_r}]
\notag \\
&\qquad 
+ \left( \frac{\nu'}{2} + \frac{\lambda'}{2}  + \frac{2}{r} \right) (T_x)\indices{^r_t}
\label{nu lambda pf eom 2 a} 
\\
0 &= \partial_t (T_x)\indices{^t_r} + \partial_r (T_x)\indices{^r_r}
+  \frac{1}{2}\left(\dot{\nu} + \dot{\lambda} \right) (T_x)\indices{^t_r}
\notag \\
&\qquad 
+ \frac{\nu'}{2} [(T_x)\indices{^r_r} - (T_x)\indices{^t_t}]
\notag \\
&\qquad 
+ \frac{1}{r} [2 (T_x)\indices{^r_r} - (T_x)\indices{^\theta_\theta} - (T_x)\indices{^\phi_\phi}],
\label{nu lambda pf eom 2 b}
\end{align}
where the first equation is for $\nu = t$ and the second is for $\nu = r$.  Moving to equilibrium variables, which requires treating everything as time-independent, Eq.\ (\ref{nu lambda pf eom 2 a}) vanishes identically and Eq.\ (\ref{nu lambda pf eom 2 b}) becomes
\begin{equation} \label{px0 eom}
p_{x0}' = - \frac{\nu'_0}{2} (\epsilon_{x0} + p_{x0}).
\end{equation}
The perturbed versions of Eqs.\ (\ref{nu lambda pf eom 2 a}) and (\ref{nu lambda pf eom 2 b}) are
\begin{align}
0 &= \delta \dot{\epsilon}_x
+ \partial_r \left[ (\epsilon_{x0} + p_{x0})v_x \right]
+ \frac{\delta \dot{\lambda}}{2}  \left( \epsilon_{x0} + p_{x0} \right)
\notag \\
&\qquad
+ \left( \frac{\nu_0'}{2} + \frac{\lambda_0'}{2} + \frac{2}{r} \right)  (\epsilon_{x0} + p_{x0})v_x
\label{per eom a}
\\
0 &=
e^{\lambda_0 - \nu_0}(\epsilon_{x0} + p_{x0}) \dot{v}_x
+ \delta p'_x + \frac{\delta \nu'}{2} (\epsilon_{x0} + p_{x0}) 
\notag\\
&\qquad
+ \frac{\nu_0'}{2} (\delta \epsilon_x + \delta p_x).
\label{per eom b}
\end{align}


\subsection{Equations for equilibrium solutions}
\label{sec:static solns}

In this subsection I put together the equations whose solutions describe equilibrium, or static, compact stars.  In doing this, I will write the equations using a different parametrization of the spherically symmetric metric than written in Eq.\ (\ref{nu lambda metric}).  I will use instead
\begin{equation}
ds^2 = - N(t,r) \sigma^2(t,r) dt^2 + \frac{dr^2}{N(t,r)} + r^2d\Omega^2,
\end{equation}
where
\begin{align}
N(t,r) &= 1 - \frac{2 G m_\text{tot}(t,r)}{r} = e^{-\lambda(t,r)}
\\
\sigma(t,r) &= e^{[\nu(t,r)+\lambda(t,r)]/2},
\end{align}
which is better suited for numerical solutions.  Dropping the time-dependence, so that the new equilibrium metric variables are $\sigma_0(r)$ and $m^\text{tot}_0(r)$, the equilibrium equations in (\ref{nu lambda equil metric eqs a}), (\ref{nu lambda equil metric eqs b}), and (\ref{px0 eom}) become
\begin{align} 
\sigma'_0 &= 4\pi G  \frac{r \sigma_0}{N_0} ( \epsilon_0^\text{tot} + p_0^\text{tot} )
\label{TOV a}
\\
m'_{x0} &= 4\pi r^2 \epsilon_{x0}
\label{TOV b}
\\
p_{x0}' &= - 
\frac{G}{r^2 N_0} \left(4\pi r^3 p_0^\text{tot} + m_0^\text{tot} \right)
(\epsilon_{x0} + p_{x0}),
\label{TOV c}
\end{align}
where
\begin{equation}
N_0 = \frac{1 - 2 G m_0^\text{tot}}{r}, \qquad
m_0^\text{tot} = \sum_x m_{x0}.
\end{equation}
Equations (\ref{TOV a})--(\ref{TOV c}) are the Tolman-Oppen\-heimer-Vol\-koff (TOV) equations adapted to a multi-fluid system \cite{Kodama} that has only gravitational inter-fluid interactions.  The individual $m_{x0}(r)$ can be interpreted as giving the total mass inside a radius $r$ for fluid $x$.

To solve the TOV equations, and thus find static compact star solutions, one specifies, say, the central values of the energy densities, $\epsilon_{x0}(0)$, from which the central values of the pressures, $p_{x0}(0)$, are obtained from the equations of state, $p_{x0}(\epsilon_{x0})$.  Equations (\ref{TOV a})--(\ref{TOV c}) can then be integrated outward from some small $r$ once appropriate inner and outer boundary conditions are determined.  Inner boundary conditions can be determined by subbing into Eqs.\ (\ref{TOV a})--(\ref{TOV c}) power law expansions of the variables, which gives $\sigma_0(0) = \sigma_c + O(r^2)$ and $m_{x0}(0) = O(r^3)$, where $\sigma_c$ is an as-yet-undetermined constant.  For outer boundary conditions, let $R_x$ be the smallest value of $r$ for which $p_{x0}(r) = 0$, which I will take to define the edge of fluid $x$.  For $r>R_x$, I set $p_{x0}(r) = 0$.  The edge of the star, $R$, is taken as the largest $R_x$.  At the edge of the star there is no longer any matter and the spacetime must match up to a Schwarzschild spacetime, and thus $\sigma_0(R) = 1$.

I have so far that the inner boundary condition $\sigma_0(0) = \sigma_c$ is as-yet-unknown, while the outer boundary condition $\sigma(R) = 1$ is known.  It is convenient to flip these by defining
\begin{equation}
\hat{\sigma}_0(r) \equiv \sigma_0(r)/\sigma_c.
\end{equation}
In using $\hat{\sigma}_0$, the only equation that changes is Eq.\ (\ref{TOV a}), which becomes
\begin{equation} \label{sigma hat eq}
\hat{\sigma}'_0 = 4\pi G  \frac{r \hat{\sigma}_0}{N_0} ( \epsilon_0^\text{tot} + p_0^\text{tot} ).
\end{equation}
The inner boundary conditions are  now $\hat{\sigma}_0(0) = 1$ and $m_{x0}(0) = 0$, and the TOV equations in (\ref{TOV b}), (\ref{TOV c}), and (\ref{sigma hat eq}) can be integrated outward from some small $r$ to the edge of the star at $r=R$, where the total mass of the star is given by
\begin{equation}
M_\text{tot} = \sum_x M_x,
\qquad
M_x = m_{x0}(R)
\end{equation}
and $\sigma_c$ is given by $\sigma_c = 1/\hat{\sigma}_0(R)$.  In this way equilibrium/static compact star solutions can be found.  I shall do this in Sec.\ \ref{sec:examples}.  

\subsection{Pulsation equations}
\label{sec:pulsation}

I now take up one of the main parts of this paper, which is the derivation of a system of pulsation equations whose solution describes radial oscillations for a system of perfect fluids with only gravitational inter-fluid interactions.  Further, the derivation is in line with Chandrasekhar's original derivation of a pulsation equation for a single fluid.  One convenience of the resulting system of pulsation equations is that the contributions from individual fluids are manifest and easily solved for, if desired.

I begin first by putting together a number of equations, which I will then use to derive the pulsation equations.  The first step is to define
\begin{equation} \label{xi}
\dot{\xi}_x \equiv v_x.
\end{equation}
Plugging this into Eq.\ (\ref{metric perturb c}) and integrating gives
\begin{equation}  \label{delta lambda eq}
\delta\lambda = 
-8\pi G r e^{\lambda_0} \sum_x (\epsilon^x_0 + p^x_0) \xi_x,
\end{equation}
which can be combined with the metric equilibrium equations (\ref{nu lambda equil metric eqs a}) and (\ref{nu lambda equil metric eqs b}) and Eq.\ (\ref{nu eq}) to obtain
\begin{align} 
&\delta \nu'(\epsilon_0^\text{tot} + p_0^\text{tot})
\label{delta nu prime}
\\
&\quad
=(\nu_0' + \lambda_0')
\left[
\delta p^\text{tot}
- \left(\nu_0' + \frac{1}{r} \right) 
\sum_x (\epsilon^x_0 + p^x_0) \xi_x
\right].
\notag
\end{align}

The equation of motion in (\ref{per eom a}), when written in terms of $\xi_x$ using Eq.\ (\ref{xi}), can immediately be integrated to give
\begin{align}
\delta \epsilon_x &= 
- \partial_r \left[(\epsilon_{x0} + p_{x0})\xi_x \right]
- \frac{\delta \lambda}{2}  \left( \epsilon_{x0} + p_{x0} \right)
\notag
\\
&\qquad - \left( \frac{\nu_0'}{2} + \frac{\lambda_0'}{2} + \frac{2}{r} \right)  (\epsilon_{x0} + p_{x0})\xi_x.
\end{align} 
Replacing $\delta \lambda$ with Eq.\ (\ref{delta lambda eq}) and then combining the result with Eqs.\  (\ref{nu lambda equil metric eqs a}) and (\ref{nu lambda equil metric eqs b}), I find
\begin{align} 
\delta \epsilon_x &=
4\pi G r e^{\lambda_0} (\epsilon_{x0} + p_{x0})
\sum_y (\epsilon_{0y} + p_{0y}) \left( \xi_y - \xi_x \right)
\notag \\
&\qquad  - \frac{1}{r^2}\partial_r \left[r^2(\epsilon_{x0} + p_{x0})\xi_x \right].
\label{delta x 1}
\end{align}

The equations of state, $p_x = p_x (\epsilon_x)$, after being perturbed and canceling their equilibrium terms, becomes $\delta p_x = (\partial p_{x0}/\partial \epsilon_{x0}) \delta \epsilon_x$.  Plugging in Eq.\ (\ref{delta x 1}), this can be written as
\begin{align}
\delta p_x 
&= 
-\xi_x 
p_{x0}'
-  p_{x0} \gamma_x
\Biggl[
\frac{ e^{\nu_0/2}}{r^2}  \partial_r \left(r^2 e^{-\nu_0/2} \xi_x \right)
\notag
\\
&\qquad - 4\pi G r e^{\lambda_0} 
\sum_y  (\epsilon_{0y} + p_{0y}) \left( \xi_y - \xi_x \right)
\Biggr],
\label{delta p eq}
\end{align}
where
\begin{equation}
\gamma_x = \left(1 + \frac{\epsilon_{x0}}{p_{x0}}\right) \frac{\partial p_{x0}}{\partial \epsilon_{x0}}
\end{equation}
is the adiabatic index for fluid $x$.

Having established the above results, I will now outline the derivation of the pulsation equations, which follow from the equation of motion in (\ref{per eom b}).  Introducing a harmonic time-dependence for all fields:
\begin{equation}
\xi_x(t,r) = \xi_x(r) e^{i\omega t},
\end{equation}
$\delta \nu (t,r) = \delta \nu(r) e^{i\omega t}$, $\delta \lambda (t,r) = \delta \lambda(r) e^{i\omega t}$, $\delta \epsilon_x (t,r) = \delta \epsilon_x(r) e^{i\omega t}$, and $\delta p_x (t,r) = \delta p_x(r) e^{i\omega t}$, where $\omega$ is the radial oscillation frequency, Eq.\ (\ref{per eom b}) becomes
\begin{align} 
&e^{-\nu_0/2} \partial_r (e^{\nu_0/2} \delta p_x)
+ \frac{\delta \nu'}{2} (\epsilon_{x0} + p_{x0}) 
+ \frac{\nu_0'}{2}\delta \epsilon_x
\notag
\\
&\qquad
=\frac{e^{\lambda_0-\nu_0/2}}{r^2}(\epsilon_{x0} + p_{x0}) \omega^2 \zeta_x,
\label{pulsation 1}
\end{align}
where I defined
\begin{equation}
\zeta_x(r) \equiv r^2 e^{-\nu_0/2} \xi_x(r).
\end{equation}
Consider first the first term in Eq.\ (\ref{pulsation 1}).  Plugging in $\delta p_x$ from Eq.\ (\ref{delta p eq}) and then using Eqs.\ (\ref{nu pp eq}) and (\ref{px0 eom}), I find
\begin{widetext}
\begin{align}
e^{-\nu_0/2} &\partial_r (e^{\nu_0/2} \delta p_x)
\notag \\
&= \frac{ e^{\nu_0/2}}{r^2}
\biggl\{
\frac{3}{r} \zeta_x p_{x0}'
- \zeta_x' p_{x0}'  
+ \frac{\nu_0'}{2}  \zeta_x \epsilon_{x0}'
+ 
\left[
8\pi G e^{\lambda_0} p_0^\text{tot}
+ \frac{\lambda_0'}{2} \left( \frac{1}{r} + \frac{\nu_0'}{2} \right)
\right] \zeta_x (\epsilon_{x0} + p_{x0}) 
\biggr\}
\notag \\
&\qquad
- e^{- \nu_0/2} \partial_r 
\left(
p_{x0} \gamma_x
 \frac{e^{\nu_0}}{r^2} \zeta_x'
\right)
+
\gamma_x p_{x0}
\frac{4\pi G}{r} e^{\lambda_0+\nu_0/2} 
\sum_y \left[(\epsilon_{0y}' + p_{0y}') \left( \zeta_y - \zeta_x \right)
+ (\epsilon_{0y} + p_{0y}) \left( \zeta_y' - \zeta_x' \right) \right]
\notag \\
&\qquad
+ 
\left[
\gamma_x
p_{x0}' 
+
\gamma_x  p_{x0} \left(- \frac{1}{r}
+ \lambda_0' + \nu_0'  \right)
+ \gamma_x' p_{x0}
\right]
\frac{4\pi G}{r} e^{\lambda_0+\nu_0/2} 
\sum_y (\epsilon_{0y} + p_{0y}) \left( \zeta_y - \zeta_x \right).
\label{pulsation a}
\end{align}
For the second term in Eq.\ (\ref{pulsation 1}), I have, after using Eqs.\ (\ref{nu lambda equil metric eqs a}), (\ref{nu lambda equil metric eqs b}), (\ref{px0 eom}), (\ref{delta nu prime}), and (\ref{delta p eq}),
\begin{align}
\frac{\delta \nu'}{2}(\epsilon_{x0} + p_{x0}) 
=
-4\pi G r
(\epsilon_{x0} + p_{x0})
\frac{e^{\nu_0/2 + \lambda_0}}{r^2} 
\sum_y
& \Biggl\{ 
\left(\frac{\nu_0'}{2} + \frac{1}{r} \right) 
(\epsilon^y_0 + p^y_0) \zeta_y
\notag \\
& + p_{y0} \gamma_y \Biggl[
\zeta_y' - 4\pi G r  e^{\lambda_0}
\sum_z (\epsilon_{0z} + p_{0z}) \left( \zeta_z - \zeta_y \right)
\Biggr] \Biggr\}.
\label{pulsation b}
\end{align}
Finally, using Eqs.\ (\ref{px0 eom}), (\ref{delta x 1}), and (\ref{pulsation b}) the second and third terms in Eq.\ (\ref{pulsation 1}) are
\begin{align}
 \frac{\delta \nu'}{2} & (\epsilon_{x0} + p_{x0}) 
+ \frac{\nu_0'}{2}\delta \epsilon_x
\notag\\
&= \frac{ e^{\nu_0/2}}{r^2}
\biggl\{
\frac{3}{r} p_{x0}'
+ 
\left[
8\pi G e^{\lambda_0} p_0^\text{tot}
+ \frac{\lambda_0'}{2} \left( \frac{1}{r} + \frac{\nu_0'}{2} \right)
\right] (\epsilon_{x0} + p_{x0}) 
\biggr\} \zeta_x
- e^{- \nu_0/2} \partial_r 
\left(
p_{x0} \gamma_x
 \frac{e^{\nu_0}}{r^2} \zeta_x'
\right)
\notag\\
&\qquad
+ 
\left[
(\gamma_x-1)
p_{x0}' 
+
\gamma_x  p_{x0} \left(- \frac{1}{r}
+ \lambda_0' + \nu_0'  \right)
+ \gamma_x' p_{x0}
\right]
\frac{4\pi G}{r} e^{\lambda_0+\nu_0/2} 
\sum_y (\epsilon_{0y} + p_{0y}) \left( \zeta_y - \zeta_x \right)
\notag\\
&\qquad +
\gamma_x p_{x0}
\frac{4\pi G}{r} e^{\lambda_0+\nu_0/2} 
\sum_y \left[(\epsilon_{0y}' + p_{0y}') \left( \zeta_y - \zeta_x \right)
+ (\epsilon_{0y} + p_{0y}) \left( \zeta_y' - \zeta_x' \right) \right].
\label{pulsation c}
\end{align}

Equations (\ref{pulsation a}) and (\ref{pulsation c}) can be plugged into Eq.\ (\ref{pulsation 1}).  The point is to construct an equation in which the only perturbation present is $\zeta_x$ and its derivatives.  In putting everything together, I will move to the metric functions $\hat{\sigma}_0(r)$, $m_{x0}(r)$, and $N_0(r)$ that were introduced in Sec.\ \ref{sec:static solns} and which are more convenient for numerical solutions.  The system of pulsation equations is
\begin{align}
&\partial_r^2 (\widehat{\Pi} \zeta_x') +(\widehat{Q}_x + \hat{\omega}^2 W_x) \hat{\zeta}_x
+ \widehat{R}
\left[ 
\left(\frac{\epsilon_{x0} + p_{x0} }{r} - p_{x0}' \right) 
\sum_y
(\epsilon_{y0} + p_{y0}) \hat{\zeta}_y
+
\frac{r^2 (\epsilon_{x0} + p_{x0})}{\hat{\sigma}_0^2 N_0}
\sum_y
\eta_y 
\right]
\notag \\
&\qquad = \widehat{S}_x \sum_y (\epsilon_{y0} + p_{y0}) \left( \hat{\zeta}_y - \hat{\zeta}_x \right)
+ 
\frac{r^2}{\hat{\sigma}_0^2 N_0}  
\widehat{R}^2
(\epsilon_{x0} + p_{x0}) 
\sum_y
\sum_z
 p_{y0} \gamma_y (\epsilon_{z0} + p_{z0}) \left( \hat{\zeta}_z - \hat{\zeta}_y \right)
\notag \\
&\qquad\qquad  +
\widehat{R} \gamma_x p_{x0}
\sum_y \left[(\epsilon_{y0}' + p_{y0}') \left( \hat{\zeta}_y - \hat{\zeta}_x \right)
+ (\epsilon_{y0} + p_{y0}) \left( \hat{\zeta}_y' - \hat{\zeta}_x' \right) \right],
\label{pulsation}
\end{align}
\end{widetext}
where
\begingroup
\allowdisplaybreaks
\begin{align}
\hat{\zeta}_x &= \sigma_c^2 \zeta_x
\notag \\
\hat{\omega} &= \omega/\sigma_c
\notag \\
\widehat{\Pi}_x &= \frac{1}{r^2}p_{x0} \gamma_x \hat{\sigma}^2 N_0
\notag \\
W_x &= \frac{1}{r^2 N_0} (\epsilon_{x0} + p_{x0} )
\notag \\
\widehat{Q}_x &=
-\frac{ \hat{\sigma}^2_0 N_0}{r^2}
\biggl\{
\frac{3}{r} p_{x0}'
+ 
 \biggl[
\frac{8\pi G}{N_0} p_0^\text{tot}
(\epsilon_{x0} + p_{x0}) 
\notag \\
&\quad
+ \left(\frac{4\pi G r}{N_0} \epsilon_0^\text{tot} - \frac{Gm_0^\text{tot}}{r^2 N_0} \right)  \left( \frac{\epsilon_{x0} + p_{x0} }{r} - p_{x0}'\right)
\biggr] 
\biggr\}
\notag \\
\widehat{R} &= 4\pi G 
\frac{\hat{\sigma}_0^2}{r}
\notag \\
\widehat{S}_x &= \widehat{R}
\biggr\{
(\gamma_x-1)
p_{x0}' 
+ \gamma_x' p_{x0}
\notag \\
&\quad +
\gamma_x  p_{x0} \bigg[ \frac{8\pi G r}{N_0} (\epsilon_0^\text{tot} + p_0^\text{tot} )  - \frac{1}{r}\biggr]
\biggr\}
\notag \\
\epsilon_{x0}' &+ p_{x0}'
= \frac{p_{x0}'}{\gamma_x}
\left( 1 + \gamma_x + \frac{\epsilon_{x0}}{p_{x0}} \right).
\label{pulsation defs}
\end{align}
\endgroup
Note that all functions with a hat are functions that have been scaled by powers of $\sigma_c = \sigma_0(0)$, as explained in Sec.\ \ref{sec:static solns}.  Importantly, we see that the only perturbation present in Eqs.\ (\ref{pulsation}) and (\ref{pulsation defs}) is $\hat{\zeta}_x$ and its derivatives.

Before discussing boundary conditions, a couple comments are in order.  The right hand side of Eq.\ (\ref{pulsation}) vanishes for a single fluid, in which case the left hand side of Eq.\ (\ref{pulsation}) is equivalent to Chandrasekhar's pulsation equation \cite{Chandrasekhar:1964zz}.  Though equivalent, the left hand side of Eq.\ (\ref{pulsation}) is not written in an identical form to Chandrasekhar's because, in the presence of multiple fluids, terms cannot cancel and combine in the same way.  When solving Eq.\ (\ref{pulsation}) I have found it best to do so in conjunction with solving the equilibrium TOV equations in (\ref{TOV b}), (\ref{TOV c}), and (\ref{sigma hat eq}).  Specifically, I have found that solving them all simultaneously is faster than first solving the TOV equations and then interpolating between points in the equilibrium solution for use in solving the pulsation equations.

To actually solve the pulsation equation I define $\eta_x \equiv \widehat{\Pi}_x \hat{\zeta}_x'$ and solve the system of first order differential equations made up of Eq.\ (\ref{pulsation}), but written in terms of $\eta_x$, and 
\begin{equation} \label{pulsation 2}
\hat{\zeta}_x'  = \frac{\eta_x}{\widehat{\Pi}_x}.
\end{equation}
Doing this requires inner and outer boundary conditions.  Inner boundary conditions are found by plugging power series expansions of the variables into the equations, which reproduce the inner boundary conditions for the equilibrium variables given in Sec.\ \ref{sec:static solns} and $\zeta_x = O(r^3)$ and $\eta_x = \eta^x_c + O(r^2)$, where the $\eta^x_c$ are as-yet-unknown constants.  A look at the equations shows that the set of $\hat{\zeta}_x$ and $\eta_x$ that solve them can all be divided by the same constant and still be solutions.  I can use this to set
\begin{equation}
\eta_1 = 1 + O(r^2),
\end{equation}
i.e.\ $\eta^{x=1}_c = 1$.  I still need to determine the $\eta^{x\neq 1}_c$.  My method for doing this is explained in the next paragraph.  The outer boundary conditions for the equilibrium variable were discussed in Sec.\ \ref{sec:static solns}.  The outer boundary conditions for the perturbations are $\eta_x(R_x) = 0$.  Intuitively this makes sense, as outside the fluid the perturbation $\zeta_x$ should stop changing.  It can be derived by requiring the Lagrangian pressure for fluid $x$ to be zero at $r = R_x$ \cite{Misner:1974qy, Chanmugam, Gondek:1997fd}.

The as-yet-undetermined constants are $\eta^{x\neq 1}_c$ and the squared radial oscillation frequency $\omega^2$.  I determine them using the shooting method.  That is, I choose values for $\omega^2$ and the $\eta^{x\neq 1}_c$.  The remaining inner boundary conditions are $\hat{\sigma}_0(0)=\eta^1_c = 1$ and $m_{x0}(0) = \zeta_x(0) = 0$.  I can then integrate the TOV equations (\ref{TOV b}), (\ref{TOV c}), and (\ref{sigma hat eq}) and the pulsation equations (\ref{pulsation}) and (\ref{pulsation 2}) outward from some small $r$.  I vary $\omega^2$ and the $\eta^{x\neq 1}_c$ until the smallest value of $r$ for which $\eta_x(r) = 0$ is exactly $r= R_x$, and hence the outer boundary conditions are satisfied.  I will have then found a solution.  If $\omega^2 < 0$, the solution is unstable to small radial perturbations.


\section{Examples}
\label{sec:examples}

In this section I present one-, two-, and three-fluid examples in which I solve for the squared radial oscillation frequency, $\omega^2$, and determine stability.  In all cases, for simplicity, I take each fluid to be a free Fermi gas.  The well-known equation of state and number density for a free Fermi gas are \cite{ShapiroBook, GlendenningBook}
\begin{align}
\epsilon_x &= 
\frac{1}{2\pi^2} 
\int_0^{k_{xF}} dk \, k^2 \sqrt{k^2 + m_{xf}^2}
\notag \\
&= \frac{1}{8\pi^2}
\Biggl[ k_{xF} \sqrt{k_{xF}^2 + m_{xf}^2}( 2k_{xF}^2 + m_{xf}^2 )
\notag \\
&\qquad - m_{xf}^4 \ln \left( \frac{k_{xF} + \sqrt{k_{xF}^2 + m_{xf}^2}}{m_{xf}} \right) \Biggr]
\\
p_x &= 
\frac{1}{6\pi^2} \int_0^{k_{xF}} dk \frac{k^4}
{\sqrt{k^2 + m_{xf}^2}}
\notag\\
&= \frac{1}{24\pi^2}
\Biggl[ k_{xF} \sqrt{k_{xF}^2 + m_{xf}^2}( 2k_{xF}^2 - 3m_{xf}^2 )
\notag \\
&\qquad + 3m_{xf}^4 \ln \left( \frac{k_{xF} + \sqrt{k_{xF}^2 + m_{xf}^2}}{m_{xf}} \right) \Biggr]
\\
n_x &= \frac{k_{xF}^3}{3\pi^2},
\end{align}
where $m_{xf}$ is the fermion mass and $k_{xF}$ is the Fermi momentum for fluid $x$.  It is convenient to define
\begin{align}
\chi_x &\equiv k_{xF} / m_{xf}
\\
t_x &\equiv 4\ln \left( \chi_x + \sqrt{1+\chi_x^2} \right),
\end{align}
so that the equation of state and number density can be written in the parametric form
\begin{align}
\epsilon_x &= \frac{m_{xf}^4}{32\pi^2}
\left( \sinh t_x - t_x \right)
\label{eps(t)}
\\
p_x &= \frac{1}{3} \frac{m_{xf}^4}{32\pi^2}
\bigl[ \sinh t_x -  8\sinh(t_x/2) + 3t_x \bigr]
\label{p(t)}
\\
n_{x}&= \frac{m_{xf}^3}{3\pi^2} \sinh^3(t_x/4).
\label{n(t)}
\end{align}
The equilibrium versions of these equations are given by simply replacing $\epsilon_x$, $p_x$, $n_{x}$ and $t_x$ with $\epsilon_{x0}$, $p_{x0}$, $n_{x0}$, and $t_{x0}$.

To find static compact star solutions, I can numerically solve the TOV equations in (\ref{TOV b}), (\ref{TOV c}), and (\ref{sigma hat eq}), as explained in Sec.\ \ref{sec:static solns}.  However, instead of evolving the pressure $p_{x0}$ using Eq.\ (\ref{TOV c}), it is much simpler to evolve $t_{x0}$ using
\begin{equation} \label{TOV t}
t_{x0}' = \frac{1}{dp_{x0}/dt_{x0}} p_{x0}',
\end{equation}
where $p_{x0}'$ is given by Eq. (\ref{TOV c}) and
\begin{equation}
\frac{dp_{x0}}{dt_{x0}} = \frac{m_{xf}^4}{12\pi^2} \sinh^4(t_{x0}/4),
\end{equation}
which follows from the equilibrium version of Eq.\ (\ref{p(t)}).

The general procedure for finding static compact star solutions and their squared radial oscillation frequencies is as described in Secs.\ \ref{sec:static solns} and \ref{sec:pulsation}.  The only changes specific to the free Fermi gas that I make is in the use of the variable $t_{x0}$:\  To find static compact star solutions, I use the central values $t_{x0}(0)$, which are related to the central values for the energy density, pressure, and number density through Eqs.\ (\ref{eps(t)})--(\ref{n(t)}).  I integrate outward from some small $r$ using the TOV equations in (\ref{TOV b}), (\ref{sigma hat eq}), and (\ref{TOV t}).  I take the edge of fluid $x$, whose location I label as $r=R_x$, to be the smallest value of $r$ such that $t_{x0}(r)=0$.  Defining the edge of the fluid this way is equivalent to defining it as $p_{x0}(r) = 0$.  For $r > R_x$, I set $t_{x0} = 0$.  The edge of the star is given by the largest $R_x$.

In presenting results in the following subsections, I will make use of the dimensionless variables
\begin{align}
\bar{r} &\equiv \frac{m_{1f}^2}{m_P} r,& \quad
\bar{\epsilon}_{x0} &\equiv \frac{4\pi}{m_{1f}^4} \epsilon_{x0},& \quad
\bar{p}_{x0} &\equiv \frac{4\pi} {m_{1f}^4} p_{x0}
\notag \\ \notag \\
\bar{\omega} &\equiv \frac{m_P}{m_{1f}^2} \omega,&
\bar{m}_{x0} &\equiv \frac{m_{1f}^2}{m_P^3} m_{x0},&
\bar{m}_{xf} &\equiv \frac{m_{xf}}{m_{1f}},
\end{align}
where $m_P = 1/\sqrt{G}$ is the Planck mass.  I have also $\bar{n}_{x0} \equiv n_{x0} / m_{1f}^3$, $\overline{M}_x \equiv (m_{1f}^2/m_P^3) M_x$, $\overline{R}_x \equiv (m_{1f}^2/m_P) R_x$, $\bar{m}_\text{tot} = \sum_x \bar{m}_x$, and $\overline{M}_\text{tot} = \sum_x \overline{M}_x$.  Regardless of the number of fluids, I choose to scale variables using $m_{1f}$, the first fluid's fermion mass.


\subsection{One fluid}

\begin{figure}
\centering
\includegraphics[width=3.1in]{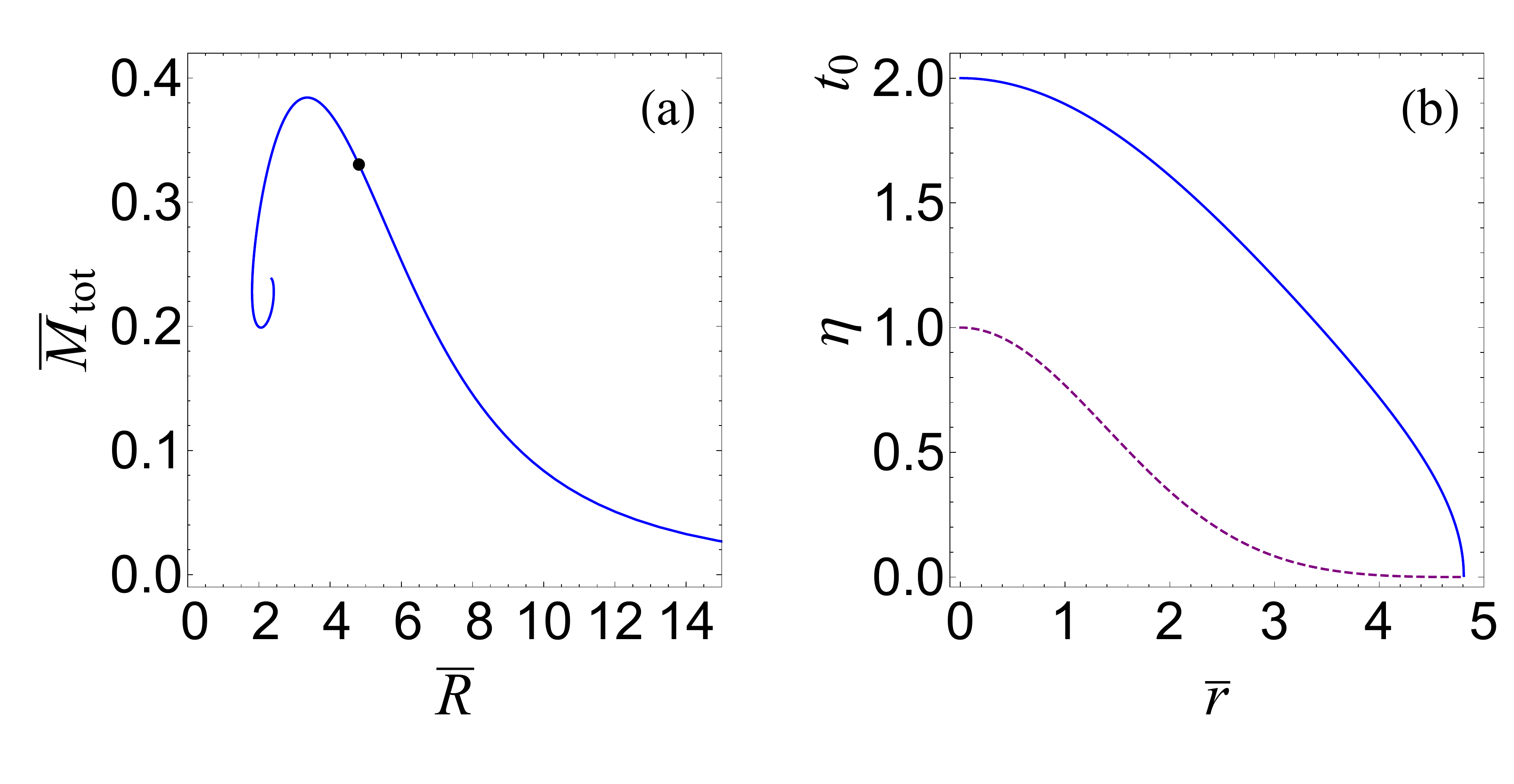}
\caption{(a) (Total) mass, $\overline{M}_\text{tot}$, versus radius, $\overline{R}$, for static/equilibrium solutions of a free Fermi gas.  (b) The solid curve plots the equilibrium solution $t_0(\bar{r})$ located at the dot in (a), with $\overline{M}_\text{tot} = 0.3303$, $\overline{R} = 4.8078$, and $t_0(0) = 2.0$.  The dashed curve plots the pulsation variable $\eta(\bar{r})$.  A solution to the pulsation equation is found when $\eta$ first equals zero at the edge of the star.}
\label{fig:1fluid}
\end{figure}

With only a single fluid I can drop the subscripted $x$s on all variables. Figure \ref{fig:1fluid}(a) shows the well-known (total) mass versus radius curve for the static/equilibrium solutions for a free Fermi gas \cite{Narain:2006kx}. Each point on the curve corresponds to a different choice for the central value $t_{0}(0)$ (or, equivalently, different choices for the central values $\epsilon_0(0)$, $p_0(0)$, or $n_0(0)$).  An important question is whether each point on the curve corresponds to an equilibrium solution that is stable or unstable with respect to small radial perturbations.

\begin{figure}
\centering
\includegraphics[width=2.35in]{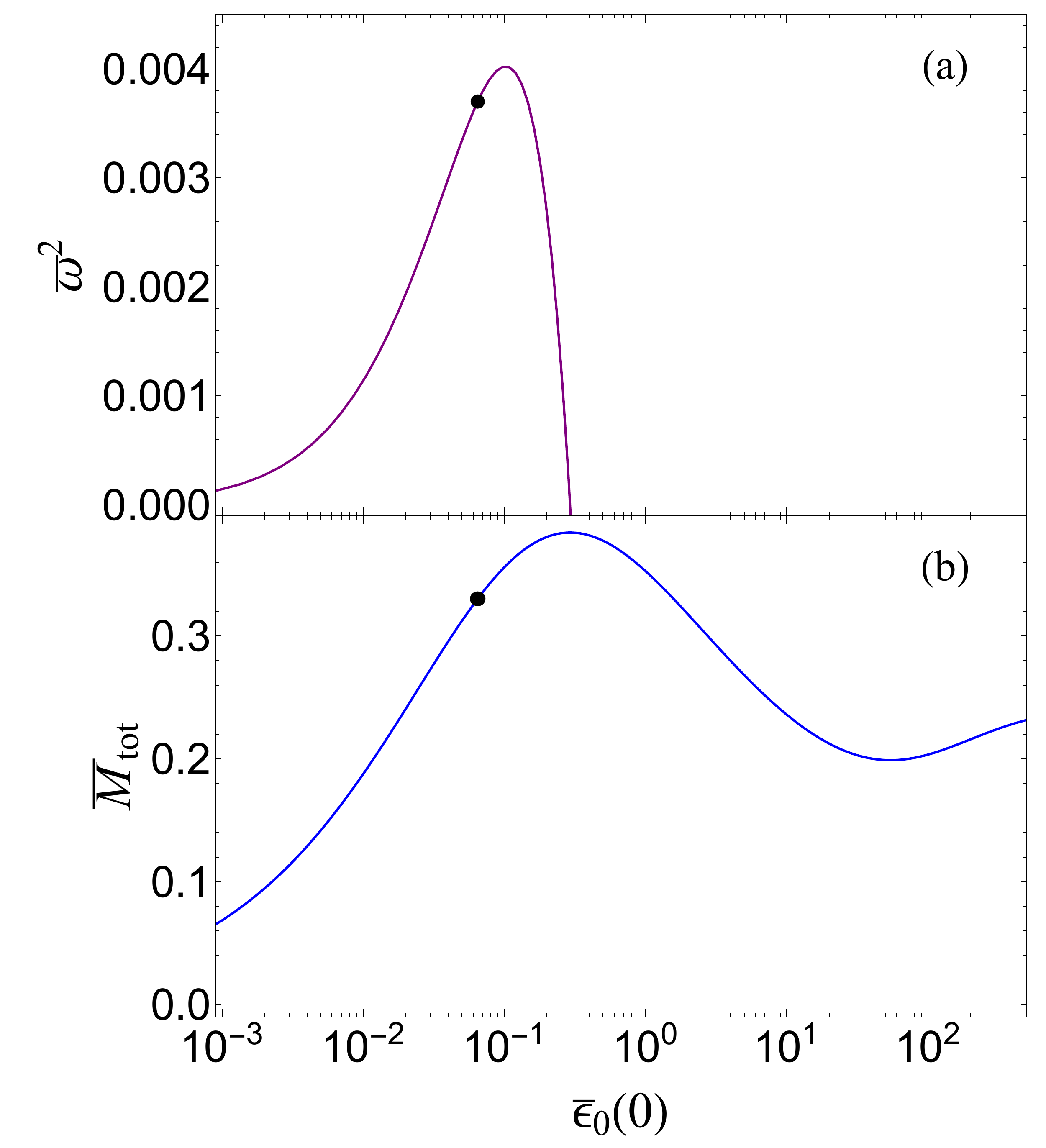}
\caption{(a) Squared radial oscillation frequency, $\bar{\omega}^2$, as a function of the central energy density, $\bar{\epsilon}_0(0)$.  (b) (Total) mass, $M_\text{tot}$, also as a function of the central energy density.  Both plots are for a free Fermi gas and the curve in (b) displays the same equilibrium solutions shown in Fig.\ \ref{fig:1fluid}(a).  The squared oscillation frequency is seen to transition from positive to negative, and hence the equilibrium solutions transition from stable to unstable, at the equilibrium solution with the largest mass.  The curves in Fig.\ \ref{fig:1fluid}(b) correspond to the solutions located at the black dots. }
\label{fig:1fluid2}
\end{figure}

Stability may be determined by computing the squared radial oscillation frequency, $\omega^2$, for the particular equilibrium solution, which is done by solving the pulsation equation.  An example of a solution to the pulsation equation is shown in Fig.\ \ref{fig:1fluid}(b).  The top solid curve plots $t_0(\bar{r})$ for the equilibrium solution indicated by the dot in Fig.\ \ref{fig:1fluid}(a).  The bottom dashed curve plots $\eta(\bar{r})$.  As explained in Sec.\ \ref{sec:pulsation}, $\eta(0) = 1$ and a solution to the pulsation equation is found when the smallest value of $\bar{r}$ for which $\eta(\bar{r}) = 0$ occurs precisely at the edge of the star at $\bar{r}=\overline{R}$.  The advantage in plotting $t_0$ in Fig.\ \ref{fig:1fluid}(b) instead of, say, $\bar{p}_0$, is that $t_0$ heads toward $t_0 = 0$ relatively abruptly at the edge of the star, making the edge of the star easily discernible in plots such as Fig.\ \ref{fig:1fluid}(b).

For a single fluid, it is well-known that the pulsation equation can be written in Sturm-Liouville form \cite{Misner:1974qy, Kokkotas:2000up}.  This tells us that a single equilibrium solution has an infinite number of radial oscillation frequencies, corresponding to an infinite number of solutions to the pulsation equation \cite{ShapiroBook}.  When the solution to the pulsation equation is found by having $r=R$ be the smallest $r$ such that $\eta(r) = 0$, the oscillation frequency found is the fundamental, which is the smallest of the oscillation frequencies.  As the oscillation frequency is increased, $\eta(r)$ develops nodes, i.e.\ additional locations where $\eta(r) = 0$.  A solution is still found when $\eta(R) = 0$, so that the outer boundary condition is satisfied.  In this paper I consider only fundamental frequencies, and thus the smallest frequencies.  

Figure \ref{fig:1fluid2}(a) plots the squared radial oscillation frequencies as a function of the central energy density $\bar{\epsilon}_0(0)$ when they are positive, and hence for the stable equilibrium solutions.  Figure \ref{fig:1fluid2}(b) plots the (total) mass of the equilibrium solutions also as a function of the central energy density $\bar{\epsilon}_0(0)$ and is an alternative to Fig.\ \ref{fig:1fluid}(a) as a way of presenting the space of equilibrium solutions.  It is well-known that for a single fluid, the transition from stable equilibrium solutions to unstable ones occurs at the equilibrium solution with the largest mass \cite{ShapiroBook, GlendenningBook}.  This is exactly what is seen in Fig.\ \ref{fig:1fluid2}, where the squared frequencies transition from positive to negative at precisely the location of the largest mass equilibrium solution.  The dots in Fig.\ \ref{fig:1fluid2} indicate the solutions shown in Fig.\ \ref{fig:1fluid}(b), just as the dot in Fig.\ \ref{fig:1fluid}(a) does.


\subsection{Two fluids}

\begin{figure*}
\centering
\includegraphics[width=6.2in]{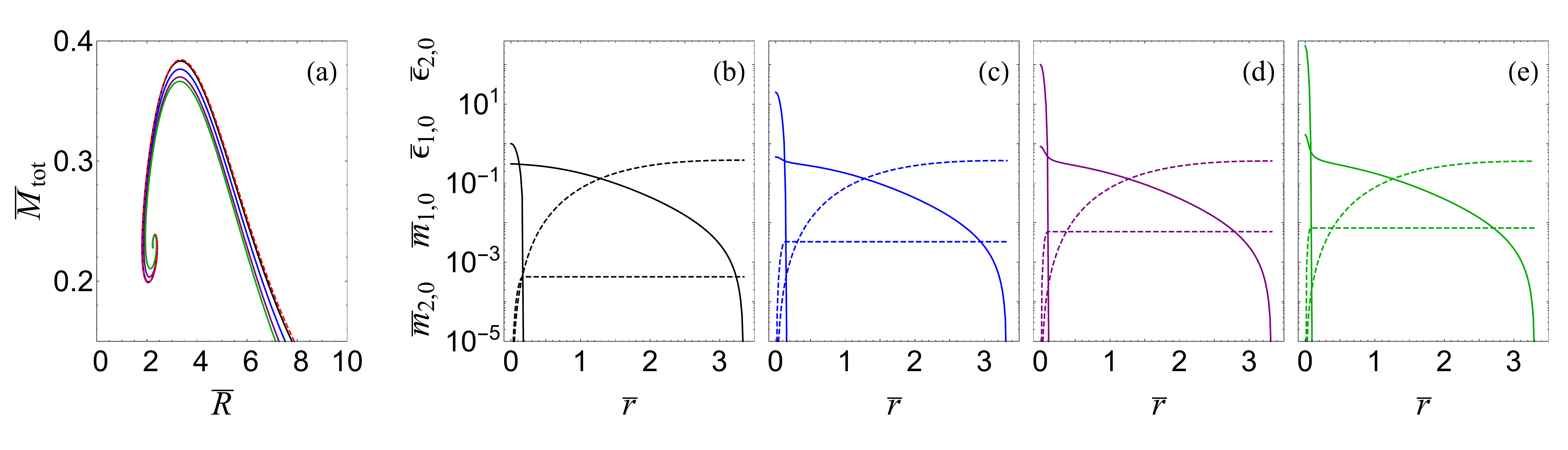}
\caption{(a) The solid lines plot the (total) mass, $\overline{M}_\text{tot}$, versus radius, $\overline{R}$, for static/equilibrium solutions for two  fluids, each of which are a free Fermi gas, with $\bar{m}_{2f} = 7$.  The dashed line is the single fluid case from Fig. \ref{fig:1fluid}(a), which is included for comparison.  The solid curves can be defined by the central energy density of fluid 2, $\bar{\epsilon}_{2,0}(0)$, which are (from top to bottom at the point of maximum $\overline{M}_\text{tot}$), 1 (black), 20 (blue), 100 (purple), and 300 (green).  (b)-(e) plot the maximum $\overline{M}_\text{tot}$ two-fluid equilibrium solutions in (a).  The solid curves plot $\bar{\epsilon}_{x0}(\bar{r})$ for both fluids (with $\epsilon_{2,0}(0) = 1$ in (b), 20 in (c), 100 in (d), and 300 in (e)).  The dashed curves plot $m_{x0}(\bar{r})$ for both fluids, which gives the total mass inside a radius $\bar{r}$ for fluid $x$.  We can see that fluid 2 dominates the energy density of the core, while fluid 1 dictates the total mass of the star (which is given by $\bar{m}_{1,0} + \bar{m}_{2,0}$ at the edge of the star) and the radius of the star (which is given by the edge of the outermost fluid).}
\label{fig:2fluid}
\end{figure*}

\begin{figure}
\centering
\includegraphics[width=2.35in]{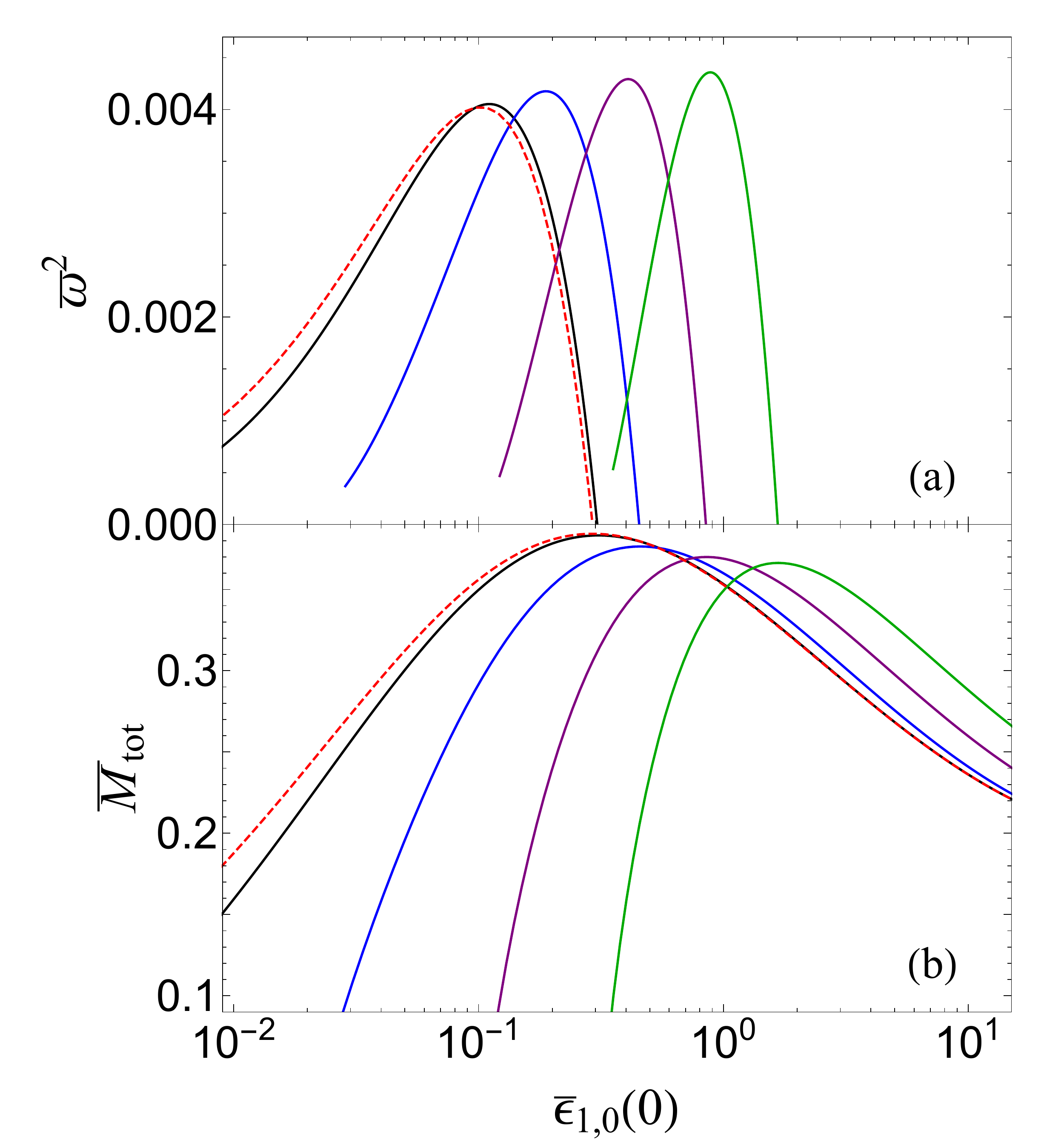}
\caption{The squared radial oscillation frequencies, $\bar{\omega}^2$, are shown in (a) for the equilibrium solutions shown in (b).  The equilibrium solutions in (b) are the same as shown in Fig.\ \ref{fig:2fluid}(a).  Just as in the single fluid case (see Fig.\ \ref{fig:1fluid2}), we see here for two fluids that the squared oscillation frequencies transition from positive to negative, and hence the equilibrium solutions transition from stable to unstable, at the equilibrium solution with the largest $\overline{M}_\text{tot}$.}
\label{fig:2fluid2}
\end{figure}

\begin{figure*}
\centering
\includegraphics[width=6.2in]{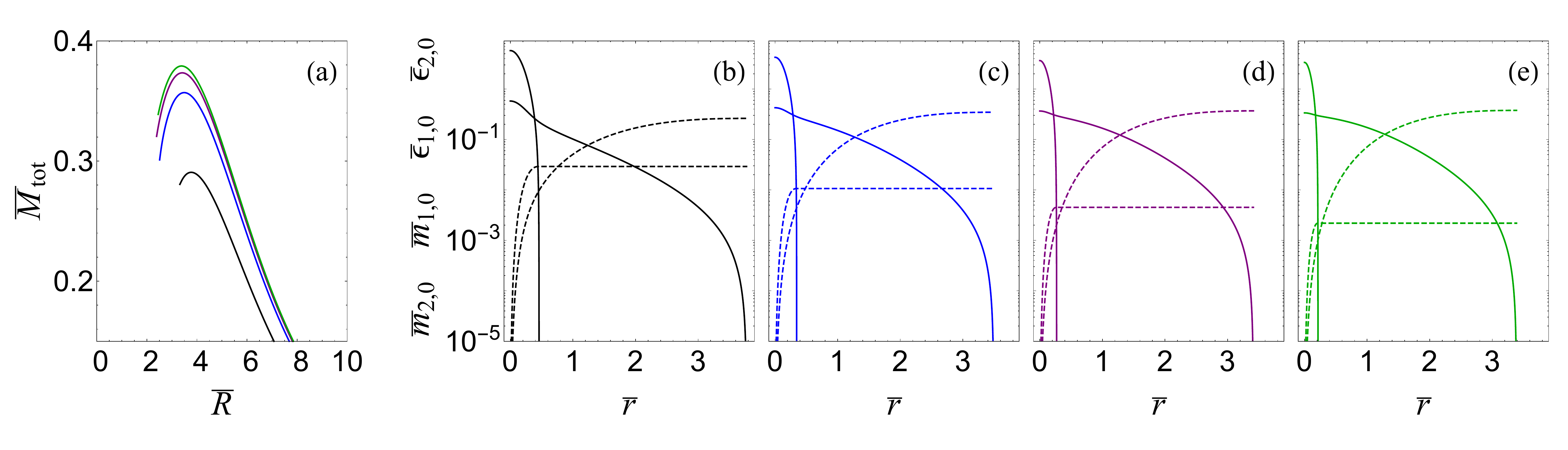}
\caption{These plots are analogous to those shown in Fig.\ \ref{fig:2fluid}.  All curves are made by fixing $\bar{\epsilon}_{2,0}(0) / \bar{\epsilon}_{1,0} = 10$.  The individual curves have (from top to bottom in (a)) $\bar{m}_{2f} = 3$ (b), 4 (c), 5 (d), and 6 (e).  In (b)--(e), the solid curves display $\bar{\epsilon}_{1,0}(r)$ and $\bar{\epsilon}_{2,0}(r)$ and the dashed curves display $\bar{m}_{1,0}(r)$ and $\bar{m}_{2,0}(r)$.}
\label{fig:2fluid3}
\end{figure*}

\begin{figure*}
\centering
\includegraphics[width=6.2in]{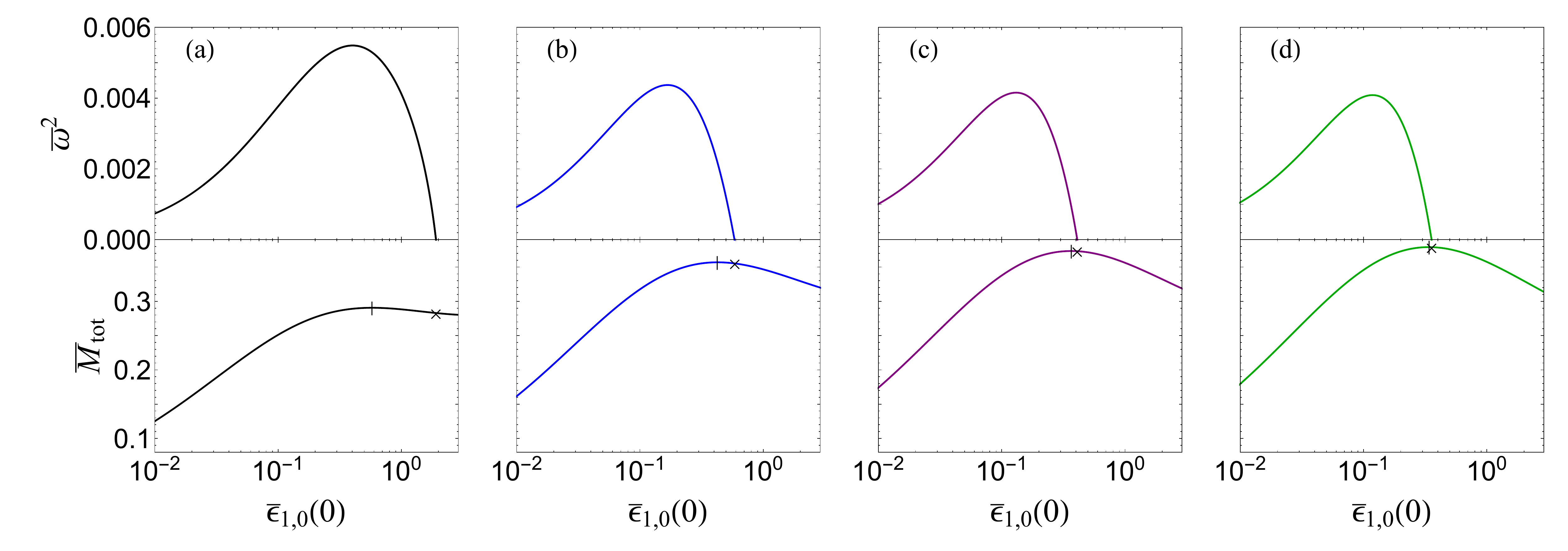}
\caption{The top plots display the squared radial oscillation frequencies and the bottom plots display the same equilibrium solutions shown in Fig.\ \ref{fig:2fluid3}(a).  The curves have $\bar{m}_{2f} = 3$ (a), 4 (b), 5 (c), and 6 (d).  In the bottom plots, the vertical lines mark the maximum mass solutions and the crosses mark the transition from stable to unstable.}
\label{fig:2fluid4}
\end{figure*}

For two fluids, the first set of results I show is for when the fermion mass of the second fluid is seven times larger than the fermion mass of the first fluid, $\bar{m}_{2f} = 7$.  I fix the central energy density of the second fluid, $\bar{\epsilon}_{2,0}(0)$, and scan through values of the central energy density of the first fluid, $\bar{\epsilon}_{1,0}(0)$.  The (total) mass versus radius curves for the resulting equilibrium/static solutions are shown in Fig.\ \ref{fig:2fluid}(a).  Each curve is for a different fixed value of $\bar{\epsilon}_{2,0}(0)$.  I have also included in this plot, for comparison, the single fluid curve from Fig.\ \ref{fig:1fluid}(a) as the dashed line.  

To get a sense of the equilibrium solutions, in Figs.\ \ref{fig:2fluid}(b)--(e) I show the maximum (total) mass equilibrium solutions from each of the two-fluid curves in Fig.\ \ref{fig:2fluid}(a).  Moving from Fig.\ \ref{fig:2fluid}(b) to Fig.\ \ref{fig:2fluid}(e) means increasing $\bar{\epsilon}_{2,0}(0)$.  The solid curves in Figs.\ \ref{fig:2fluid}(b)--(e) display $\bar{\epsilon}_{1,0}(\bar{r})$ and $\bar{\epsilon}_{2,0}(\bar{r})$ and the dashed curves display $\bar{m}_{1,0}(\bar{r})$ and $\bar{m}_{2,0}(\bar{r})$, which gives the mass of fluid $x$ inside a radius $\bar{r}$.  In Figs.\ \ref{fig:2fluid}(b)--(e) we can can see that fluid 2 dominates the core of the star, while fluid 1 dictates the total mass ($\overline{M}_\text{tot} = \bar{m}_{1,0}(\overline{R}) + \bar{m}_{2,0}(\overline{R}))$ and radius of the star.

Figure \ref{fig:2fluid2}(a) shows the squared radial oscillation frequencies for the equilibrium solutions shown in Fig.\ \ref{fig:2fluid2}(b) (which are the same equilibrium solutions shown in Fig.\ \ref{fig:2fluid}(a)).  In both figures, the curves are given as functions of the first fluid's central energy density, $\bar{\epsilon}_{1,0}(0)$.  I have only plotted the oscillation frequencies and equilibrium solutions for when fluid 1 affects $\overline{M}_\text{tot}$.  This is the reason why the oscillation frequency curves in Fig.\ \ref{fig:2fluid2}(a) begin at the same value of $\bar{\epsilon}_{1,0}(0)$ as the equilibrium solution curves do in  Fig.\ \ref{fig:2fluid2}(b).  Just as in the single-fluid case, we see that the squared frequencies transition from positive to negative, and hence the equilibrium solutions transition from stable to unstable, at the equilibrium solution with the largest mass.

I now present a different set of results.  This time I fix $\bar{\epsilon}_{2,0}(0) / \bar{\epsilon}_{1,0}(0) = 10$ and find solutions for $\bar{m}_{2f} = 3$, 4, 5, and 6.  Figure \ref{fig:2fluid3}(a) displays the (total) mass versus radius curves and Figs.\  \ref{fig:2fluid3}(b)--(e) displays the maximum (total) mass equilibrium solutions, analogously to Fig.\ \ref{fig:2fluid}.

Figure \ref{fig:2fluid4} displays the squared radial oscillation frequencies (in the top plots) for the same equilibrium solutions (shown in the bottom plots) as displayed in \ref{fig:2fluid3}(a).  In the bottom plots, the vertical lines mark the maximum mass solutions and the crosses mark the transition from stable to unstable.  One can see that the transition from stable to unstable does not occur at the equilibrium solution with the largest mass.  This is not unexpected, and, in general, occurs in multiple-component stars (see, for example, \cite{Henriques:1990xg, ValdezAlvarado:2012xc}).


\subsection{Three fluids}

\begin{figure}
\centering
\includegraphics[width=2in]{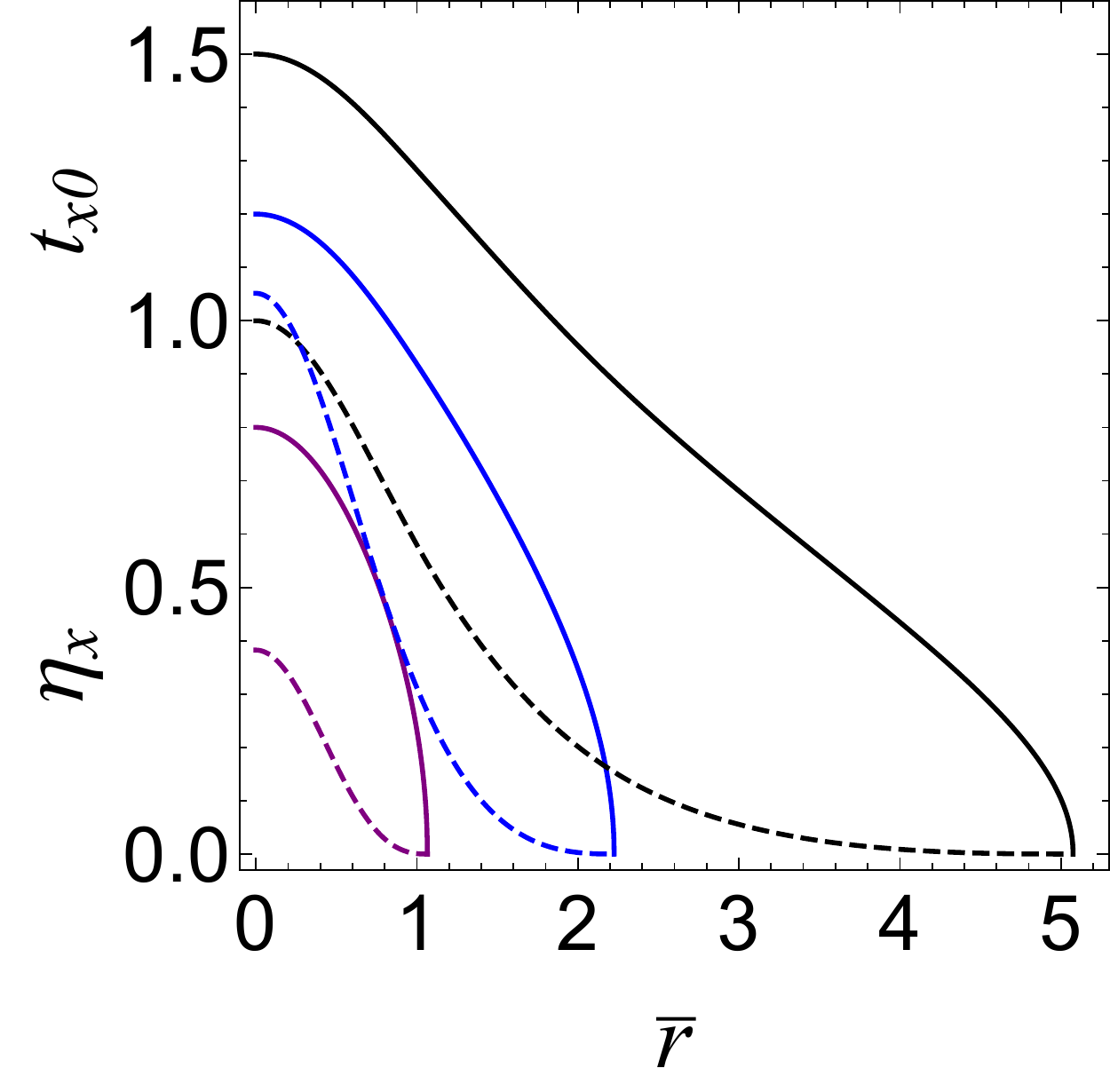}
\caption{Three-fluid solution, analogous to the one-fluid solution shown in Fig.\ \ref{fig:1fluid}(b).  The fermion masses for the fluids are $\bar{m}_{2f} = 1.5$ and $\bar{m}_{3f} = 2$.  The solid curves plot the equilibrium variables $t_{x0}(\bar{r})$, with (from top to bottom) $t_{1,0}(0) = 1.5$ (black), $t_{2,0}(0) = 1.2$ (blue), and $t_{3,0}(0) = 0.8$ (purple).  The dashed curves plot the pulsation variables $\eta_x(\bar{r})$.  This solution has $\overline{M}_\text{tot} = 0.1072$, $\overline{R} = 5.0756$, and $\bar{\omega}^2 = 0.003091$ and is stable.}
\label{fig:3fluid}
\end{figure}

With three fluids there are three parameters to be determined using the shooting method:\ $\bar{\omega}^2$, $\eta_2(0)$, and $\eta_3(0)$.  It is significantly more time consuming to determine three parameters than it is to determine two parameters and I present results for only a single equilibrium solution in Fig.\ \ref{fig:3fluid}.  Figure \ref{fig:3fluid} is analogous to Fig.\ \ref{fig:1fluid}(b), with the solid curves plotting the equilibrium variables $t_{x0}(\bar{r})$ and the dashed curves plotting the pulsation variables $\eta_x(\bar{r})$.


\section{Conclusion}
\label{sec:conclusion}

In this work I derived a system of pulsation equations for spherically symmetric, multiple-fluid compact stars, under the assumption that the only inter-fluid interactions are gravitational.  The solution to the pulsation equations is the squared radial oscillation frequency for the star.  I solved the system of pulsation equations in one-, two-, and three-fluid examples and used the squared radial oscillation frequency to determine stability with respect to small perturbations.  As far as I am aware, this is the first time that a pulsation equation has been solved for a three-fluid compact star.

\vfill



%

\end{document}